\documentclass{elsart}
\usepackage{amsfonts}
\usepackage{amssymb}
\usepackage{amsmath}

\input{tcilatex}

\begin{document}

\begin{center}
\bigskip \bigskip \textsl{\ }{\Large Stock market volatility: An approach
based on Tsallis entropy}

\bigskip
\end{center}

\begin{description}
\item \textbf{S\'{o}nia R. Bentes}$^{1\ast }$, \textbf{Rui Menezes}$^{2}$, 
\textbf{Diana A. Mendes}$^{2}$

$^{1}$ISCAL, Av. Miguel Bombarda, 20, 1069-035 Lisboa, Portugal

$^{2}$ISCTE, Av. Forcas Armadas, 1649-025 Lisboa, Portugal.
\end{description}

\begin{center}
\bigskip

\bigskip

{\large \textbf{Abstract}}
\end{center}

\QTP{Body Math}
\bigskip $\bigskip $

One of the major issues studied in finance that has always intrigued, both
scholars and practitioners, and to which no unified theory has yet been
discovered, is the reason why prices move over time. Since there are several
well-known traditional techniques in the literature to measure stock market
volatility, a central point in this debate that constitutes the actual scope
of this paper is to compare this common approach in which we discuss such
popular techniques as the standard deviation and an innovative methodology
based on Econophysics. In our study, we use the concept of Tsallis entropy
to capture the nature of volatility. More precisely, what we want to find
out is if Tsallis entropy is able to detect volatility in stock market
indexes and to compare its values with the ones obtained from the standard
deviation. Also, we shall mention that one of the advantages of this new
methodology is its ability to capture nonlinear dynamics. For our purpose,
we shall basically focus on the behaviour of stock market indexes and
consider the CAC $40$, MIB $30$, NIKKEI $225$, PSI $20$, IBEX $35$, FTSE $%
100 $ and SP $500$ for a comparative analysis between the approaches
mentioned above.

\bigskip

\bigskip

\textbf{PACS (2008):} 87.23.Ge, 89.65.Gh, 89.70.Cf, 89.90.+n$\bigskip $


$\bigskip $

$^{\ast }$E-mail: soniabentes@clix.pt

\section*{Introduction}

In the last few years there has been an increasing debate on the subject of
stock market volatility. In spite of its present relevance, this is not an
entirely new issue and has emerged in a systematic way when Shiller \cite%
{Journal-1} first argued that the observed stock market volatility was
inconsistent with the predictions of the present value models, quite popular
in the past. Moreover, Grossman and Shiller \cite{Journal-2} found out that
the intemporal variation appeared to be inexplicably high and could not be
rationalized even in models with a stochastic discount factor. Even though
some authors questioned the conclusion of excessive volatility, like Flavin 
\cite{Journal-3} or Kleidon \cite{Journal-4}, latter tests accounting for
dividend nonstationarity and small sample bias continued to lend support to
Shiller's initial claim (see Refs. \cite{Journal-5}, \cite{Journal-6}, \cite%
{Journal-7}, \cite{Journal-8}, \cite{Journal-9}). A new insight into this
was brought by Schwert \cite{Journal-10}, who asked the seminal question
"Why does stock market volatility change over time?", having reached the
conclusion that only a small amount of fluctuations could be explained by
models of stock valuation. In this light, many other studies have appeared
with the aim of studying every single aspect of stock market volatility,
giving rise to an intense debate on the theme. Recognizing its relevance,
Daly \cite{Journal-11} summarizes some of the major reasons pointed out for
its study: (i) Firstly, when market exhibits an excess volatility, investors
may find it difficult to explain it based only upon the information about
the fundamental economic factors. As a result an erosion of confidence and a
reduced flow of capital into equity markets may occur. (ii) Secondly, for
firms individually considered, volatility is an important factor in
determining the probability of bankruptcy. The higher the volatility for a
given capital structure, the higher the probability of default. (iii)
Thirdly, volatility is also an important factor in determining the bid-ask
spread. So, the higher the volatility of the stock the wider will be the
spread between bid and ask prices, thus affecting the market liquidity. (iv)
Fourthly, hedging techniques such as portfolio insurance are affected by the
volatility level, with the prices of insurance increasing with volatility.
(v) Fifthly, if consumers are risk averse, as the financial theory suggests,
an increase in volatility will therefore imply a reduction in economic
activity with adverse consequences for investment. (vi) Finally, increased
volatility over time may induce regulatory agencies and providers of capital
to force firms to allocate a larger percentage of available capital to cash
equivalent investments, to the potential detriment of allocation efficiency.

In this brief overview we have tried to shed some light on the theme and to
unfold some of its major implications. Nevertheless, given the
impracticability of analyzing volatility as a whole we focus on its
particular aspect of measurement. Here, however we face an obstacle: since
volatility is not observed, there has been no agreement on how to measure
it, thus emerging a plethora of techniques. Another conclusion that appeared
to have arisen is that volatility is volatile.

The main contribution of this paper is to compare two different approaches:
one based on the statistical measure of the standard deviation or variance
and another centred on the concept of entropy. In this regard, we
particularly focus on the concept of Tsallis entropy, which constitutes a
generalization of the Boltzmann-Gibbs or Shannon entropy. These measures
were both generated in the domain of physics, although the latter is also
attributed to the Information Theory, and their application to financial
phenomena falls in the domain of the so-called econophysics. In an analogy
with terms like biophysics, geophysics and astrophysics this word was
originally introduced by Stanley \textit{et al.} \cite{Journal-12} in an
attempt to legitimize the study of economics by physicists. One argument is
that some regularities were found between these two areas. Another argument
points out the benefits of the experimental method commonly used in physics,
which departs from the observed data without imposing any previous model.
Also, it is worthy to note the evidence of common research interests between
these two areas. As Mantegna and Stanley \cite{Book-1} pointed out, an
active domain of research in physics is the characterization of prices
changes, \textit{i.e.}, volatility. In our particular research we apply the
concept of entropy to capture the presence of nonlinear dynamics in seven
international stock market indexes since the standard deviation can only
detect linear relationships. The empirical analysis is conducted with data
from Datastream in order to perform a comparative research.

The remainder of the paper is organized as follows: Section 1 describes the
most commonly used measure of volatility - the standard deviation - and
compares it with two different measures of entropy: the Shannon entropy and
its generalization - the Tsallis entropy. Section 2 exhibits the empirical
findings, and Section 3 draws the conclusions.

\section{Volatility and Entropy Measures: Some Concepts}

In this Section we define various measures of volatility. We begin with the
standard deviation and then analyze the Tsallis entropy and its special
case: the Shannon entropy. Before proceeding further on we shall first
clarify the term volatility. According to a wide range of research,
volatility can be broadly defined as the changeableness of the variable
under consideration (see \cite{Journal-1} and \cite{Journal-13} for some
references). As a result, the more this variable fluctuates over time, the
more volatile that variable is said to be. Usually, this term is popular as
a synonymous of risk and uncertainty; though its meaning is not quite the
same. Yet, they are related concepts. Knight \cite{Book-2} established the
difference between both of them in the following sense: while in a situation
of risk we are not certain about the results of a given action but know
exactly its probability distribution function in uncertainty the \textit{%
p.d.f.} is always unknown.

Another view was introduced by Hwang and Satchell \cite{Journal-14}, who
considered that volatility could be regarded as a combination of two
components: transitory noise and permanent fundamental volatility. While the
former is temporary and caused by the trading noise, the latter is generated
by the arrival of information. This is in accordance with the work of Ross 
\cite{Journal-15}, who has already pointed out the role of information in
this context.

Based on the fact that volatility could be not constant over time, \textit{%
i.e.}, "volatility is volatile", some authors have divided the various
techniques in two different categories: time invariant (or independent) and
time variant (or dependent) measures. In the first group we include the
techniques studied in this paper, since they are time independent. The other
one clearly exceeds the scope of our research and is related to, for
example, the ARCH (Autoregressive Conditionally Heteroskedastic) models, and
their subsequent derivations.

\subsection{\protect\bigskip \textbf{A traditional measure of volatility}}

A popular way of measuring volatility is to compute the returns $R_{t}$ of
the asset under consideration

\begin{equation}
R_{t}=\ln P_{t}-\ln P_{t-1},  \label{log-returns}
\end{equation}%
where $P_{t}$ and $P_{t-1}$ denote the prices at time $t$ and $t-1$,
respectively, and then estimate the corresponding standard deviation over
some historical period $T$.

\begin{equation}
\sigma =\sqrt{\frac{\sum_{1}^{T}\left( R_{t}-\overline{R}\right) ^{2}}{T-1}},
\label{SD}
\end{equation}%
with $\overline{R}$ representing the sample average return, $\overline{R}%
=\sum R_{t}/T$.

Although this measure has some advantages since it is simple to estimate and
has the ability to capture the probability of occurring extreme events, it
also shows some drawbacks. One is that it could lead to an abrupt change in
volatility once shocks fall out of the measurement sample. And, if shocks
are still included in a relatively long measurement sample period, then an
abnormally large observation will imply that the forecast will remain in an
artificial high level even though the market is subsequently tranquil.
Secondly, it assumes that recent and more distant events are equally
weighted. However, the most likely situation is that the more recent ones
have a stronger effect on volatility than the older ones. Finally, it only
captures linear relationships, ignoring all kinds of nonlinear dynamics
among data. In this light, some more sophisticated measures have emerged
aiming to improve the understanding of volatility. With regard to this, a
measure that appears to be particularly relevant is the concept of entropy,
which constitutes our major aim in this study.

Nonetheless, it is worthy to note that, in spite of all the flaws that have
been recognized by a wide body of research, the standard deviation is still
the most popular measure of volatility being used as a benchmark for
comparing the forecast ability of more complex models.

\subsection{\textbf{Entropy as a measure of volatility}}

An alternative way to study stock market volatility is by applying concepts
of physics which significant literature has already proven to be helpful in
describing financial and economic phenomena. One measure that can be applied
to describe the nonlinear dynamics of volatility is the concept of entropy.
This concept was originally introduced in 1865 by Clausius to explain the
tendency of temperature, pressure, density and chemical gradients to flatten
out and gradually disappear over time. Based on this, Clausius developed the
Second Law of Thermodynamics which postulates that the entropy of an
isolated system tends to increase continuously until it reaches its
equilibrium state. Although there are many different understandings of this
concept, the most commonly used in literature is as a measure of ignorance,
disorder, uncertainty or even lack of information (see \cite{Journal-16}).
Later, in a subsequent investigation, Shannon \cite{Journal-17} provided a
new insight into this matter showing that entropy was not only restricted to
thermodynamics but could instead be applied in any context where
probabilities can be defined. In fact, thermodynamic entropy can be viewed
as a special case of the Shannon entropy since it measures probabilities in
the full state space. Based on the Hartley's \cite{Journal-18} formula,
Shannon derived his entropy measure and established the foundations of
information theory.

For the probability distribution $p_{i}\equiv p\left( X=i\right) $, $\left(
i=1,...,n\right) $ of a given random variable $X,$ Shannon (Boltzmann-Gibbs)
entropy $S(X)$ for the discrete case, can be defined as 
\begin{equation}
S\left( X\right) =-\sum\limits_{i=1}^{n}p_{i}\ln p_{i},
\end{equation}%
with the conventions $0\ln \left( 0/z\right) =0$ for $z\geq 0$ and $z\ln
\left( z/0\right) =\infty $.

As a measure of uncertainty the properties of entropy are well established
in literature (see \cite{Book-3}). For the non-trivial case where the
probability of an event is less than one, the logarithm is negative and the
entropy has a positive sign. If the system only generates one event, there
is no uncertainty and the entropy is equal to zero. By the same token, as
the number of likely events duplicates the entropy increases one unit.
Similarly, it attains its maximum value when all likely events have the same
probability of occurrence. On the other hand, the entropy of a continuous
random variable may be negative. The scale of measurements sets an arbitrary
zero corresponding to a uniform distribution over a unit volume. A
distribution which is more confined than this has less entropy and will be
negative.

Shannon entropy has been most successful in the treatment of equilibrium
systems in which short/space/temporal interactions dominate. However, there
are many anomalous systems in nature that just do not verify the simplifying
assumption of ergodicity and independence. Some examples are:
metaequilibrium states in large systems involving long range forces between
particles; metaequilibrium states in small systems ($100-200$ particles);
glassy systems; some classes of dissipative systems, mesoscorpic systems
with nonmarkovian memory. With the aim of studying this kind of systems,
Tsallis \cite{Journal-19} derived a generalized form of entropy, known as
Tsallis entropy. Although this measure was first introduced by Havrda and
Charv\'{a}t \cite{Journal-20} in cybernetics and later improved by Dar\'{o}%
czy \cite{Journal-21}, it was Tsallis \cite{Journal-19} who really developed
it in the context of physical statistics and, therefore, it is also known as
Havrda-Charv\'{a}t-Dar\'{o}czy-Tsallis entropy.

For any nonnegative real number $q$ and considering the probability
distribution $p_{i}\equiv p$ $(X=i)$, $i=1,...,n$ of a given random variable 
$X,$ Tsallis entropy denoted by $S_{q}\left( X\right) $ for the discrete
case, is defined as 
\begin{equation}
S_{q}\left( X\right) =\frac{1-\sum\limits_{i=1}^{n}p_{i}^{q}}{q-1}
\end{equation}%
where the $q-$\textit{exponential} function is defined by

\begin{equation}
y=\left[ 1+\left( 1-q\right) x\right] ^{\frac{1}{1-q}}\equiv e_{q}^{x}\qquad
\left( e_{1}^{x}=e^{x}\right)
\end{equation}%
whose inverse is the $q-$\textit{logarithm} function

\begin{equation}
\ln _{q}x\equiv \frac{x^{1-q}-1}{1-q}\qquad \left( \ln _{1}x=\ln x\right) 
\text{.}
\end{equation}

The entropic index $q$ characterizes the statistics we are dealing with; as $%
q\rightarrow 1$, $S_{q}\left( X\right) $ recovers $S\left( X\right) $ since
the $q$-logarithm uniformly converges to a natural logarithm as $%
q\rightarrow 1$. This index may be regarded as a biasing parameter since $%
q<1 $ privileges rare events and $q>1$ privileges common events (see \cite%
{Journal-22}). A concrete consequence of this is that while the Shannon
entropy yields exponential equilibrium distributions, Tsallis entropy yields
power-law distributions. As Tatsuaki and Takeshi \cite{Journal-23} have
already pointed out, the index $q$ plays a similar role as the light
velocity $c$ in special relativity or Planck's constant $\hbar $ in quantum
mechanics in the sense of a one-parameter extension of classical mechanics,
but unlike $c$ or $\hbar $, $q$ does not seem to be a universal constant.
Further, we shall mention that for applications of finite variance $q$ must
lie within the range $1\leq q<5/3$. Additionally, in the case of financial
series Tsallis, Anteneodo, Borland and Osorio \cite{Journal-22} have proven
that $q\simeq 1.4-1.5$.

Tsallis entropy exhibits a series of notable properties described as follows
(see, for example, Refs. \cite{Journal-24}, \cite{Arxiv-1}, \cite{Journal-25}%
, \cite{Journal-19}):

i) Non-negativity: $S_{q}\left( X\right) \geq 0$ for any arbitrary set $%
\left\{ p_{i}\right\} $. The equality holds for $q>0$ and certainty (all
probabilities equal zero excepting one which equals unity).

ii) Equiprobability: If $p_{i}=1/W$, $\forall _{i}$ (microcanonical
ensemble) we obtain, $\forall _{q}$, the following \textit{extreme} value:

\begin{equation}
S_{q}\left( X\right) =\frac{W^{1-q}-1}{1-q}.
\end{equation}

iii) Pseudo-additivity: If $A$ and $B$ are two independent systems (\textit{%
i.e.}, $p_{ij}^{A+B}=p_{i}^{A}+p_{j}^{B}$), we verify that

\begin{equation}
\frac{S_{q}\left( A+B\right) }{k}=\frac{S_{q}\left( A\right) }{k}+\frac{%
S_{q}\left( B\right) }{k}+\left( 1-q\right) \frac{S_{q}\left( A\right) }{k}%
\frac{S_{q}\left( B\right) }{k},
\end{equation}

since in all cases $S_{q}\left( X\right) \geq 0$, $q<1$, $q=1$ and $q>1$
respectively correspond to \textit{superadditivity} (\textit{%
supreextensivity - }$S_{q}\left( A+B\right) >$\textit{\ }$S_{q}\left(
A\right) +S_{q}\left( B\right) $), \textit{additivity} (\textit{extensivity
- }$S_{q}\left( A+B\right) =$\textit{\ }$S_{q}\left( A\right) +S_{q}\left(
B\right) $) \ and \textit{subadditivity} (\textit{subextensivity - }$%
S_{q}\left( A+B\right) <$\textit{\ }$S_{q}\left( A\right) +S_{q}\left(
B\right) $).

iv) Additivity: It has been recently shown that $S_{q}\left( X\right) $ is
also extensive, \textit{i.e.},

\begin{equation}
S_{q}\left( A_{1}+A_{2}+...+A_{N}\right) \simeq
\sum\limits_{i=1}^{n}S_{q}\left( A_{i}\right) ,
\end{equation}

for special kinds of correlated systems, more precisely when the phase-space
is occupied in a scale-invariant form (see, \cite{Journal-26}, \cite%
{Journal-27}, for some references). By being extensive for an appropriate
value of $q$, $S_{q}\left( X\right) $ complies with Clausius' concept of
macroscopic entropy and with thermodynamics.

v) Reaction under bias: The Shannon entropy can be rewritten as

\begin{equation}
S\left( X\right) =-\left[ \frac{d}{dx}\sum\limits_{i=1}^{W}p_{i}^{x}\right]
_{x=1}.
\end{equation}

This can be seen as a reaction to a translation of the bias $x$ in the same
way as differentiation can be seen as a reaction of a function under a
(small) \textit{translation} of the abscissa. Along the same line, $%
S_{q}\left( X\right) $ can be rewritten as

\begin{equation}
S_{q}\left( X\right) =-\left[ D_{q}\sum\limits_{i=1}^{W}p_{i}^{x}\right]
_{x=1},
\end{equation}%
where

\begin{equation}
D_{q}h\left( X\right) \equiv \frac{h\left( qx\right) -h\left( x\right) }{qx-x%
}\qquad \left( D_{1}h\left( x\right) =\frac{dh\left( x\right) }{dx}\right)
\end{equation}%
is Jackson's 1909 generalized derivative, which can be seen as a reaction of
a function under \textit{dilatation} of the abscissa (or under a \textit{%
finite} increment of the abscissa).

vi) Concavity: If we consider two probability distributions $\left\{
p_{i}\right\} $\ and $\left\{ p_{i}^{\prime }\right\} $ for a given system $%
\left( i=1,...,W\right) $, we can define the convex sum of the two
probability distributions as

\begin{equation}
p_{i}^{\prime \prime }\equiv \mu p_{i}+\left( 1-\mu \right) p_{i}^{\prime
}\qquad \qquad \left( 0<\mu <1\right) \text{.}
\end{equation}

An entropic functional $S\left( \left\{ p_{i}\right\} \right) $ is said 
\textit{concave} if and only if for all $\mu $ and for all $\left\{
p_{i}\right\} $\ and $\left\{ p_{i}^{\prime }\right\} $

\begin{equation}
S\left( \left\{ p_{i}^{\prime \prime }\right\} \right) \geq \mu S\left(
\left\{ p_{i}\right\} \right) +\left( 1-\mu \right) S\left( \left\{
p_{i}^{\prime }\right\} \right) \text{.}
\end{equation}

By concavity we mean the same property where $\geq $ is replaced by $\leq $.
It can be shown that the entropy $S_{q}\left( X\right) $ is concave (convex)
for every $\left\{ p_{i}\right\} $ and every $q>0$ $\left( q<0\right) $. It
is important to stress that this property implies, in the framework of
statistical mechanics, thermodynamic stability, \textit{i.e.}, stability of
the system with regard to energetic perturbations.

vii) Stability or experimental robustness: An entropic functional $S\left(
\left\{ p_{i}\right\} \right) $ is said to be \textit{stable} or \textit{%
experimentally robust} if and only if, for any given $\varepsilon >0$,
exists $\delta _{\varepsilon }>0$ such that, independently from $W$,

\begin{equation}
\sum\limits_{i=1}^{W}\left\vert p_{i}-p_{i}^{\prime }\right\vert \leq \delta
_{\varepsilon }\implies \left\vert \frac{S\left( \left\{ p_{i}\right\}
\right) -S\left( \left\{ p_{i}^{\prime }\right\} \right) }{S_{\max }}%
\right\vert <\varepsilon \text{.}
\end{equation}

This implies that

\begin{equation}
\lim_{\varepsilon \rightarrow 0}\lim_{W\rightarrow \infty }\left\vert \frac{%
S\left( \left\{ p_{i}\right\} \right) -S\left( \left\{ p_{i}^{\prime
}\right\} \right) }{S_{\max }}\right\vert =\lim_{W\rightarrow \infty
}\lim_{\varepsilon \rightarrow 0}\left\vert \frac{S\left( \left\{
p_{i}\right\} \right) -S\left( \left\{ p_{i}^{\prime }\right\} \right) }{%
S_{\max }}\right\vert =0\text{.}
\end{equation}

Lesche \cite{Journal-28} has argued that the experimental robustness is a
necessary requisite for an entropic functional to be a physical quantity
because it essentially assures that, under arbitrary small variations of the
probabilities, the relative variation of entropy remains small.

Since its proposal, Tsallis entropy has been the source of most empirical
research devoted not exclusively to physics but also comprising other
scientific areas such as biology, chemistry, geophysics, medicine, economics
and finance. It is our aim in this study to especially address the latter
and find out whether Tsallis entropy is useful to measure stock market
volatility.

\section{Empirical Results}

This Section explores the empirical relevance of the theoretical results
obtained by both perspectives. To do so we have gathered\ data from several
countries in order to detect whether some similarities can be found among
them. This is especially relevant in the context of the globalization we are
living in, which also constitutes another area of research interest.

\subsection{ Data}

In our empirical research the data set compounds the daily returns of the
CAC $40$ (France), MIB $30$ (Italy), NIKKEI $225$ (Japan), PSI $20$
(Portugal), IBEX $35$ (Spain), FTSE $100$ (U.K) and SP $500$ (U.S.A.)
extending from $8$ January $1990$ to $7$ April $2006$. Each index contains $%
4240$ observations, which is large enough to make our analysis meaningful.
These data were collected on a daily basis without considering the
re-investment of dividends and were computed in accordance with Eq. \ref%
{log-returns} where the closing prices were the inputs. Fig. 1 plots the
results.\bigskip \bigskip

\FRAME{dhFU}{5.4345in}{4.6112in}{0pt}{\Qcb{Fig. 1 Data plot of the index
returns in the period 8 January 1990 to 7 April 2006}}{}{Figure}{\special%
{language "Scientific Word";type "GRAPHIC";maintain-aspect-ratio
TRUE;display "USEDEF";valid_file "T";width 5.4345in;height 4.6112in;depth
0pt;original-width 8.0211in;original-height 6.7991in;cropleft "0";croptop
"1";cropright "1";cropbottom "0";tempfilename
'K7NNNZ01.wmf';tempfile-properties "PR";}}

\bigskip

As a preliminary analysis we may say that all indexes show evidence of
changing volatility. However, a more in-depth analysis is required to draw
consistent conclusions to this regard, as performed in the next subsections.
Table 1 presents some descriptive statistics.

{\scriptsize \bigskip Table 1 Summary statistics of the daily returns }

{\scriptsize \noindent 
\begin{tabular}{lccccccc}
\hline
Statistics & CAC 40 & MIB 30 & NIKKEI 225 & PSI 20 & IBEX 35 & FTSE 100 & SP
500 \\ \hline\hline
Mean & $0.000287$ & $0.000215$ & $-7.67E-05$ & $0.000199$ & $0.000352$ & $%
0.000246$ & $0.000341$ \\ 
Median & $0.000220$ & $0.000000$ & $0.000000$ & $0.000000$ & $0.000308$ & $%
0.000235$ & $0.000257$ \\ 
Maximum & $0.061670$ & $0.069017$ & $0.093935$ & $0.062732$ & $0.063311$ & $%
0.054147$ & $0.053666$ \\ 
Minimum & $-0.073584$ & $-0.077873$ & $-0.072108$ & $-0.080149$ & $-0.082164$
& $-0.053676$ & $-0.070264$ \\ 
Skewness & $-0.177881$ & $-0.182288$ & $0.058954$ & $-0.405571$ & $-0.299988$
& $-0.158930$ & $-0.126998$ \\ 
Kurtosis & $6.413817$ & $6.078641$ & $6.857375$ & $11.48285$ & $6.853787$ & $%
6.536921$ & $7.295682$ \\ 
Jarque-Bera & $2081.259$ & $1967.934$ & $2631.14$ & $12828.95$ & $2687.391$
& $2227.915$ & $3271.407$ \\ 
Probability & $0.000000$ & $0.000000$ & $0.000000$ & $0.000000$ & $0.000000$
& $0.000000$ & $0.000000$ \\ \hline
\end{tabular}
}

{\scriptsize \bigskip }

From a statistical point of a view there is evidence of weak negative
asymmetry in all the returns considered, excluding the NIKKEI 225, which
presents a weak positive asymmetry (together with a negative mean). In
addition, all indexes exhibit excess kurtosis. As a consequence,
unconditional normality is significantly rejected as the Jarque-Bera test $p$%
-value is less than 0.01 in all cases. In this light, there is strong
evidence of fat-tails for all series, as expected.

\subsection{Standard Deviation Results}

We now proceed to the analysis of the standard deviation results as depicted
in Fig. 2.\bigskip \bigskip \bigskip \FRAME{dtbpFU}{5.1854in}{2.7622in}{0pt}{%
\Qcb{Fig. 2 Relative standard deviation of the stock indexes returns}}{}{%
Figure}{\special{language "Scientific Word";type
"GRAPHIC";maintain-aspect-ratio TRUE;display "USEDEF";valid_file "T";width
5.1854in;height 2.7622in;depth 0pt;original-width 5.7934in;original-height
3.0727in;cropleft "0";croptop "1";cropright "1";cropbottom "0";tempfilename
'K7SKJU00.wmf';tempfile-properties "PR";}}\bigskip

The standard deviation is a measure of dispersion of a probability
distribution that depends on the value of the underlying mean. Therefore, a
more convenient representation of the volatility is based on the coefficient
of variation which is a normalized measure of dispersion, and, thus, it is a
dimensionless number. This measure is particularly useful for variables that
are always positive and have a positive mean so that an appropriate
alternative is the Relative Standard Deviation (RSD). The Relative Standard
Deviation is just the absolute value of the ratio of the standard deviation
to the mean multiplied by $100$. It provides a good picture of the overall
linear dispersion underlying the data. Our results show that the NIKKEI $225$
presents, by far, the highest linear volatility value among the seven
indexes under consideration. This is obviously not surprising since the
Japanese stock market was subjected to a severe instability over the period
analyzed, showing a non-increasing long-run trend in the raw price series
and quite sharp oscillations over time. This was thus transmitted to the
returns series and translates into abnormally large oscillations or high
volatility, as observed. Next, but by far lower than in the previous case,
the MIB $30$ shows the second highest value of linear volatility, followed
by Portuguese PSI $20$ and the French CAC $40$. The Spanish IBEX $35$ and
the north-American SP $500$ exhibit the lowest values of linear volatility
as measured by the Relative Standard Deviation.

In order to have an idea of the relative discrepancy of the linear
volatility coefficient across the seven markets under analysis, and taking
the north-American SP $500$ as our basis, we can observe that the Spanish
IBEX $35$ coefficient is $12\%$ higher, whereas the British FTSE $100$, the
French CAC $40$ and the Portuguese PSI $20$ are, respectively, $28\%$, $37\%$
and $49\%$ higher than the SP $500$. The MIB $30$ and the NIKKEI $225$
multiply by two and five, respectively, the SP $500$ coefficient. Stock
market volatility, therefore, appears to have a quite different pattern of
behaviour around the world when measured in a linear way. Is this just a
systematically linear behaviour or volatility also shows signs of nonlinear
dynamics across markets? This is what we shall analyze in the next section.

\subsection{Entropy Results}

\bigskip \bigskip In the domain of the econophysics approach we have
computed the Tsallis and Shannon entropies, which are depicted in Table 2.

{\scriptsize \bigskip Table 2 Shannon and Tsallis entropies }

{\scriptsize \noindent 
\begin{tabular}{lcccccccc}
\hline
Statistics & Index ($q$) & CAC 40 & MIB 30 & NIKKEI 225 & PSI 20 & IBEX 35 & 
FTSE 100 & SP 500 \\ \hline\hline
Shannon & 1 & 3.0655 & 3.073 & 2.9163 & 2.6515 & 2.8951 & 3.0644 & 2.989 \\ 
& 1.4 & 1.7229 & 1.7255 & 1.6718 & 1.5708 & 1.6977 & 1.7229 & 1.6948 \\ 
Tsallis & 1.45 & 1.6216 & 1.6238 & 1.5766 & 1.4869 & 1.5997 & 1.6217 & 1.5967
\\ 
& 1.5 & 1.5295 & 1.5313 & 1.4898 & 1.41 & 1.5104 & 1.5297 & 1.5074 \\ \hline
\end{tabular}
}

{\scriptsize \bigskip }

All entropies were estimated with histograms based on equidistant cells. For
the calculation of Tsallis entropy we have set values at $1.4$, $1.45$ and $%
1.5$ for the index $q$, which is consistent with the finding that when
considering financial data their values lie within the range $q\simeq
1.4-1.5 $ (see \cite{Journal-22}). Since all entropies are positive we shall
conclude that the data show nonlinearities. This phenomenon is more obvious
for the MIB $30$, CAC $40$ and FTSE $100$, and a little less so for the SP $%
500$, NIKKEI $225$ and PSI $20$. When we look at the relative discrepancies
of the data taking as our basis the SP $500$, the overall difference across
the seven markets is not very marked. For the Shannon entropy there is, in
most cases, a relative change of $2.5\%-3\%$, positive or negative. The
exception is the PSI $20$ that exhibits a negative change of $11\%$ relative
to the SP $500$. On the other hand, for the Tsallis entropy (using, for
example, the results for $q=1.45$), the relative change is even smaller:
around $7\%$ less for the PSI $20$ and $1\%$ for all other indexes.

Since the entropy is designed to capture the overall linear and nonlinear
dispersion (or volatility) observed in the data, our results point to the
conclusion that volatility appears to show a relatively homogeneous pattern
across international stock markets. Curiously, the \textquotedblleft
less\textquotedblright\ volatile market, the PSI $20$, is also the smallest
and the most dependent of the seven markets analyzed. Globally, the
Portuguese stock market appears to be $7\%-11\%$ less volatile than the
north-American one. However, in terms of linear dispersion, the former is $%
49\%$ more volatile than the latter. That is, the proportion of linear
volatility on the overall volatility is higher in the Portuguese market than
in the north-American one. This appears to reveal that the volatility in the
Portuguese stock market is more linearly predictable than the volatility in
the north-American market.

A similar situation occurs in the case of the Japanese stock market
relatively to the north-American one, taken as a benchmark. Here, however,
the proportion of the overall volatility explained by linear dependencies
appears to be higher than in the previous case. For all other markets, the
overall and the linear volatility figures are higher than the north-American
benchmark so that it is not possible to make any final conclusion about the
relative weight of the linear and nonlinear dispersion relatively to the US.
Apparently, however, they are all more linearly predictable than the
north-American stock market.

\section{Conclusions}

In this paper we have investigated the volatility of seven indexes: CAC $40$%
, MIB $30$, NIKKEI $225$, PSI $20$, IBEX $35$, FTSE $100$ and SP $500$. Our
major goal was to compare two different perspectives: one based on the
standard deviation and another supported by the concept of entropy. For our
purpose two variants of this notion were regarded: the Tsallis and Shannon
statistics.

In particular, the results from both entropies have shown nonlinear dynamics
in the volatility of all indexes and must be understood in complementarity.
The results, however, must be compared with the relative standard deviation
ones in order to have a full picture of the overall phenomenon. This is
especially relevant for the decision making process in which all the
information is regarded as necessary and useful. Nonetheless, in spite of
all the divergences encountered, there is an apparent common behaviour in
most European Markets.

In this study we especially address the concept of entropy as an alternative
to the standard deviation since it can capture the uncertainty and disorder
in a time series without imposing any constraints on the theoretical
probability distribution, which constitutes its major advantage.

\end{document}